\documentclass[aps,prl,twocolumn,groupedaddress]{revtex4-1}
\usepackage{amsmath,amssymb}
\usepackage{graphicx}
\usepackage{fancybox}
\usepackage{subfigure}
\usepackage{color}

\DeclareMathOperator{\erf}{erf}

\graphicspath{{./Figures/}}

\newcommand{\vect}[1]{\mathbf{#1}}

\newcommand{\vdiff}[2]{\left|\vect{#1} - \vect{#2}\right|}

\newcommand{\phiR}{\ensuremath{\phi_{\rm R}}}

\newcommand{\V}{\ensuremath{\mathcal{V}}}
\newcommand{\Vr}{\ensuremath{\mathcal{V}_{\rm R}}}

\newcommand{\vs}{\ensuremath{v_0(r)}}
\newcommand{\vl}{\ensuremath{v_1(r)}}

\newcommand{\rhoq}{\ensuremath{\rho^q}}

\newcommand{\allpos}{\vect{\overline{R}}}

\begin{document}

\title{Efficient solutions of self-consistent mean field equations for dewetting and electrostatics in nonuniform liquids}


\author{Zhonghan Hu}
\altaffiliation[Present address: ]{State Key Laboratory of Supramolecular Structure and Materials, Jilin
University, Changchun 130012, China}

\author{John D. Weeks}
\affiliation{Institute for Physical Science and Technology and
Department of Chemistry and Biochemistry, University of
Maryland, College Park, Maryland 20742}


\date{\today}

\begin{abstract}

We use a new configuration-based version
of linear response theory to efficiently solve
self-consistent mean field equations relating an effective single particle potential
to the induced density. The versatility and accuracy of the method is illustrated
by applications to dewetting of a hard sphere solute in a Lennard-Jones fluid, the interplay
between local hydrogen bond structure and electrostatics for
water confined between two hydrophobic walls, and to ion pairing in ionic solutions.
Simulation time has been reduced by more than an order of magnitude over previous methods.

\end{abstract}

\maketitle

Mean field theories have long provided a physically suggestive and qualitatively useful description of 
structure, thermodynamics, and phase transitions in condensed matter systems.
In  these approaches certain long-ranged components of the intermolecular interactions are replaced by an 
effective single particle potential or  ``molecular field'' that
depends self-consistently on the nonuniform density the field
itself induces \cite{Kadanoff:2000}.
Recent work has shown that this method can produce exceptionally 
accurate results for models of simple and molecular liquids, ionic fluids, and water
provided that i) all the intermolecular interactions averaged over are slowly varying at typical nearest 
neighbor distances where strong local forces dominate and ii) an accurate description of the density 
induced by the effective field is
used \cite{Weeks1998,Weeks2002,Chen_Weeks2004,
Rodgers_Weeks2008}.
When both these conditions are satisfied we call the resulting 
approach Local Molecular Field (LMF) theory.

LMF theory can be viewed as a mapping that relates the
structure and thermodynamics of the full long-ranged system
to those of a simpler ``mimic system'' with truncated intermolecular interactions
but in an effective or restructured field
that accounts for the averaged effects of the long-ranged
interactions \cite{Weeks1998,Weeks2002}.  Analyzing
the short-ranged mimic system may offer particular
advantages for systems with Coulomb interactions,
since standard particle-mesh Ewald sum treatments
do not scale well in massively-parallel simulations \cite{Schulz2009}.
But to  realize the full potential of LMF theory
as a practical tool in computer simulations and in qualitative analysis, 
we must efficiently determine the effective field.
We show here that when condition i) above is satisfied we can
use a new configuration-based version of linear response theory
to satisfy condition ii) as well. This leads to a highly accurate
and efficient way to solve the self-consistent LMF equation. 
These ideas may also
help solve self-consistent-field equations that appear in many other contexts.

To illustrate the method in its simplest form,
we first study dewetting of a single hard sphere ``solute'' particle
in a Lennard-Jones (LJ) fluid.
However LMF theory provides a unified
perspective and only minor changes are
needed to describe more complex systems with Coulomb interactions. Here
we focus on the coupling between dewetting and electrostatics in water confined by hydrophobic walls
and on ion pairing in ionic solutions. Long-ranged forces play a key but subtle role in
all these examples and require a very accurate solution of the LMF equation.

Consider a nonuniform LJ fluid in the presence of the field $\phi_{0}(\vect{r})$
generated by the hard sphere solute, as described below.
The LJ pair potential is separated by the sign of the force into a short-ranged
repulsive core and a longer-ranged perturbation part \cite{Weeks:1971}:
$u_{LJ}({r}_{ij})= u_{0}({r}_{ij}) + u_{1}({r}_{ij})$.
$u_{1}$ contains only slowly-varying attractive forces and
thus is suitable for LMF averaging. The LJ mimic system
has truncated intermolecular interactions
$u_{0}({r}_{ij})$ in the presence of an effective or restructured field
$\phi_{\rm R}(\vect{r})$, chosen in principle so that the nonuniform density
$\rho_{\rm R}(\vect{r};[\phiR]) \equiv \left< \rho(\vect{r},\allpos)\right>_{\phiR}$ in the mimic system (indicated
by the subscript R) equals
the density $\rho(\vect{r};[\phi_{0}])$ in the full system \cite{Weeks1998,Weeks2002,Chen_Weeks2004,
Rodgers_Weeks2008}.
Here $\left< \,\,\right>_{\phiR}$ denotes a normalized ensemble average
in the mimic system in the presence of $\phiR$,
$\allpos \equiv \{ \vect{r}_{i} \} $ denotes a microscopic configuration of all particles, and 
$  \rho(\vect{r},\allpos) \equiv  \sum_{i=1}^{N}
 \delta ( \vect{r} - \vect{r}_{i} )$
is the microscopic configurational density.

As shown in detail 
elsewhere \cite{Weeks1998,Weeks2002,Chen_Weeks2004,
Rodgers_Weeks2008},
$\phiR$ and the associated equilibrium density $\rho_{\rm R}(\vect{r};[\phiR])$
can be accurately determined by solving the self-consistent LMF equation
\begin{equation}
  \phiR(\vect{r}) = \phi_{0}(\vect{r}) + \int d\vect{r}^\prime 
  \rho_{\rm R}(\vect{r}^\prime;[\phiR])
  u_1\left(\left|\vect{r}-\vect{r}^\prime \right|\right) + C,
  \label{eqn:LMFGeneral}
\end{equation}
where $C$ is a constant setting the zero of energy \cite{generalfield}.

In earlier work \cite{Chen_Weeks2004,Rodgers_Weeks2008},
the LMF equation was solved by
straightforward iteration, using computer simulations to accurately determine
the density induced by the given field at each iteration.
However, even with a good initial estimate $\tilde{\phi}_{0}$ for the effective field,
determining the density induced by the remaining
changes in the field $\tilde{\phi}_{\rm R1}\equiv \phi_{\rm R} - \tilde{\phi}_{0}$
from further iterations required more simulations.
By rewriting the density
in Eq.\ (\ref{eqn:LMFGeneral}) so that the dependence on
$\tilde{\phi}_{\rm R1}$ appears inside the ensemble average,
we show here that the LMF equation can be iterated to self-consistency with no
new simulations required.

This is very easy to do formally. We refer to the system with effective field $\tilde{\phi}_{0}$
as a ``trial'' system and note that the total potential energy associated with the correction
$\tilde{\phi}_{\rm R1}$ in a configuration
$\allpos$ is given by
\begin{equation}
\tilde{\Phi}_{\rm R1} (\allpos)\equiv \sum_{i=1}^{N}\tilde{\phi}_{\rm R1}(\vect{r}_{i})=\int d\vect{r} \rho(\vect{r},\allpos) \tilde{\phi}_{\rm R1}(\vect{r}).
\label{eqn:Phitilde}
\end{equation}
Similarly defining $U_{0}(\allpos)$ as the total intermolecular potential energy in configuration
$\allpos$ we have exactly
\begin{align}
  \left< \rho(\vect{r},\allpos)\right>_{\phiR} &= \frac{\int d\allpos \rho(\vect{r},\allpos)
  e^{ -\beta [U_{0}(\allpos) +
 \tilde{ \Phi}_{0}(\allpos)+\tilde{\Phi}_{\rm R1}(\allpos)]}}{\int d\allpos e^{ -\beta [U_{0}(\allpos) +
  \tilde{\Phi}_{0}(\allpos)+\tilde{\Phi}_{\rm R1}(\allpos)]}} \nonumber \\
  &=  \frac{ \left< \rho(\vect{r},\allpos)e^{ -\beta \tilde{\Phi}_{\rm R1}(\allpos)}\right>_{\tilde{\phi} _{0}}}{ \left< e^{ -\beta \tilde{\Phi}_{\rm R1}(\allpos)} \right>_{\tilde{\phi} _{0}}}.
  \label{eqn:rhoexact}
\end{align}

The idea behind Eq.\ (\ref{eqn:rhoexact}), usually with a  configuration dependent function $F(\allpos)$
replacing $\rho(\vect{r},\allpos)$ and different choices of trial and
full systems, is very well known and serves as the basis for perturbation 
and weighted histogram methods for thermodynamic properties \cite{Chipot2007}. Here we
use it to determine changes of the density in the LMF equation
as the iteration proceeds to self-consistency. 

Averaging over exponentials as in Eq.\ (\ref{eqn:rhoexact}) is usually problematic,
and in most applications sophisticated techniques like umbrella
sampling are needed to obtain good statistics \cite{Chipot2007}. However,
condition i), required for the quantitative validity of LMF theory itself, ensures that only
slowly varying intermolecular forces appear in $\tilde{\Phi}_{\rm R1}(\allpos)$.
This suggests it may not be difficult to find a trial field $\tilde{\phi}_{0}$
for which the exponential remainder $e^{ -\beta \tilde{\Phi}_{\rm R1}(\allpos)}$
varies sufficiently slowly over most relevant configurations $\allpos$
that a simple direct average is accurate.

A linearization of Eq.\ (\ref{eqn:rhoexact}) provides a way to test for this condition.
Defining $\delta \tilde{\Phi}_{\rm R1}(\allpos) \equiv \tilde{\Phi}_{\rm R1}(\allpos)-
\big < \tilde{\Phi}_{\rm R1}(\allpos) \big >_{\tilde{\phi}_{0}}$ and
$\delta \rho(\vect{r},\allpos) \equiv \rho(\vect{r},\allpos) - \big < \rho(\vect{r},\allpos) \big >_{\tilde{\phi}_{0}}$,
we find
\begin{equation}
 \left< \rho(\vect{r},\allpos)\right>_{\phiR}\simeq \left< \rho(\vect{r},\allpos)\right>_{\tilde{\phi}_{0}}
 -\beta  \left< \delta \rho(\vect{r},\allpos) \delta \tilde{\Phi}_{\rm R1} (\allpos) \right>_{\tilde{\phi}_{0}}.
 \label{eqn:linearPhi}
\end{equation}
This is equivalent to the usual linear response formula
\begin{align}
\left< \rho(\vect{r},\allpos)\right>_{\phiR} &\simeq \left< \rho(\vect{r},\allpos)\right>_{\tilde{\phi}_{0}}
\nonumber \\
&-\beta   \int d \vect{r^\prime}\left< \delta \rho(\vect{r},\allpos) \delta \rho(\vect{r^\prime},\allpos) \right>_{\tilde{\phi}_{0}}
 \tilde{\phi}_{\rm R1}(\vect{r^\prime})
 \label{eqn:linres}
\end{align}
on using Eq.\ (\ref{eqn:Phitilde}).
However the linear response (LR) function 
$\left< \delta \rho(\vect{r},\allpos) \delta \rho(\vect{r^\prime},\allpos) \right>_{\tilde{\phi}_{0}}$
depends on $\vect{r}$ and $\vect{r^\prime}$ separately in a nonuniform system
and is usually too complicated to
be determined directly.
In contrast, the averages in Eqs.\ (\ref{eqn:linearPhi}) and (\ref{eqn:rhoexact})
simply reweight the bin histograms used to determine the nonuniform density
and depend on $\vect{r}$ alone. They can be easily calculated
using the saved configurations of a single well-chosen trial system.

For very slowly varying $\tilde{\Phi}_{\rm R1} (\allpos)$,
the system is in a linear regime where essentially identical
structural changes arise from the
exponential (EXP) form in Eq.\ (\ref{eqn:rhoexact}) or the LR form in Eq.\ (\ref{eqn:linearPhi}).
By requiring that a trial system gives the same result on iterating the LMF equation using both forms,
we have a conservative and objective test for
an accurate solution. We refer to this as the LR-LMF method,
since in general Eq.\ (\ref{eqn:linearPhi}) proves most useful.
To our knowledge, the advantages of the configuration-based LR formula
in Eq.\ (\ref{eqn:linearPhi}) have not been exploited before.

We first study a hard sphere solute in a LJ fluid at a state near
the triple point \cite{Huang_Chandler2000}.
The solute is represented by an external field $\phi_{0}(r)$ that is infinite inside
a cavity of radius $R_{c}$ and zero otherwise. 
In the simplest ``strong coupling approximation'' (SCA) to LMF theory, often successfully used
in perturbation theories of uniform liquids \cite{Weeks:1971}, 
all effects of the slowly varying $u_{1}$ on the molecular structure are ignored
and the effective field $\phi_{\rm R}(r)$ is approximated by the bare hard sphere field $\phi_{0}(r)$.
When $R_{c}$ (measured in units of $\sigma_{LJ}$) is unity,
attractive forces nearly cancel and the SCA gives good results, correctly predicting
an oscillatory density response with a large density maximum at contact.

However, as the cavity radius increases, particles near the cavity
experience unbalanced attractive forces from LJ particles further away that
reduce the contact density. 
%
%
\begin{figure}[tb]
  \begin{center}
    \includegraphics[width=6.0cm,angle=270]{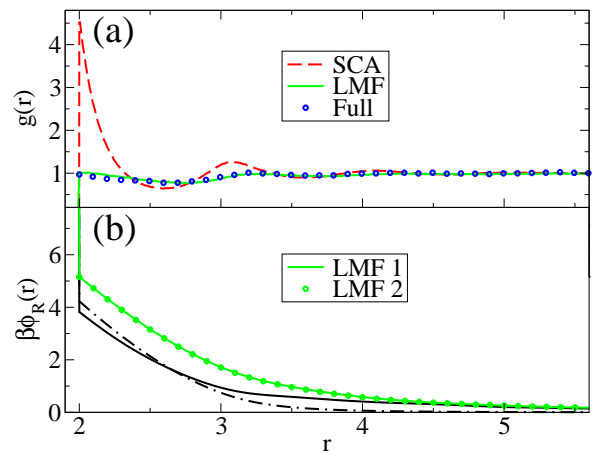}
  \end{center}
 \caption{LMF/LR/EXP treatment of the drying transition induced by a hard sphere
solute in a LJ fluid for $T=0.85$, $\rho^B=0.70$. (a)
The RDF around a hard sphere cavity
with $R_{c}=2$ for the full LJ fluid, the SCA trial system, and the
converged LMF system.
(b) Trial fields in the linear regime (solid and dot-dashed curves)
and the final self-consistent field (circles and solid line) \cite{sim-details}.}
  \label{fig:hardrying}
\end{figure}
Strong reduction already is seen for $R_{c} =2$.
The circles in Fig.\ 1a give results of computer simulations \cite{sim-details}
for the radial distribution function (RDF)
$g(r;[\phi_{0}]) \equiv \rho (r;[\phi_{0}])/ \rho^{B}$
induced by the hard sphere solute in the full LJ fluid with bulk density $ \rho^{B}$.
This exhibits an essentially structureless profile with a contact value of unity.
The solid curve in Fig.\ 1a gives $g_{\rm R}(r;[\phi_{\rm R}])$, the RDF 
in the mimic system.
The excellent agreement between the LMF prediction
and the full LJ $g(r;[\phi_{0}])$ shows that LMF theory quantitatively captures the drying effect.
This contrasts with the dashed line in Fig.\ 1a,
the density in the truncated system induced by the bare field $\phi_{0}$.
The failure of this SCA prediction
illustrates the general need in most nonuniform systems for
a proper self-consistent solution of the LMF equation.

The self-consistent field $\phi_{\rm R}(r)$ was obtained from solving
Eq.\ (\ref{eqn:LMFGeneral}) using
two different trial fields shown in Fig.\ 1b.
Both trial systems are in the linear regime and give the same final result.
The solid curve results from a previous trial simulation 
based on the SCA that used the bare $\phi_{0}(r)$.
As Fig.\ 1a suggests, this represents a very poor initial guess,
and the EXP form (\ref{eqn:rhoexact}) produced very noisy
data, indicating poor overlap between the SCA and final LMF configurations. However
Eq.\ (\ref{eqn:linearPhi}) gave much smoother data and its use
in the LMF equation gave the solid curve
in Fig.\ 1b as the output field. Using this as a second trial field generates
a converged solution of the LMF equation using either
Eqs.\ (\ref{eqn:linearPhi}) or (\ref{eqn:rhoexact}).

Since $u_{1}$ is slowly-varying, using relatively crude approximations to the asymmetric density
in the LMF equation  can often give a better trial field than the simple SCA.
Thus in Fig.\ 1b, the dot-dashed trial field was
found by approximating the density by
a step function that vanishes inside the cavity and equals $\rho^{B}$ outside.

LMF theory proves even more useful \cite{Chen_Weeks2004,
Rodgers_Weeks2008}
when applied to systems with Coulomb interactions
in the presence of an electrostatic potential  $\V(\vect{r})$ arising from a fixed external charge
distribution $\rhoq_{\rm ext}(\vect{r^\prime})$.
The basic Coulomb interaction $ 1/r \equiv \vs + \vl$ is separated into short- and long-ranged components, 
where $v_1(r)$ is proportional to the electrostatic potential
arising from a normalized Gaussian charge distribution with
width $\sigma$,
\begin{equation}
  \label{eqn:v1def}
 v_1(r) \equiv  \frac{1}{\pi^{3/2}\sigma^3}\int e^{-r^{\prime 2}/\sigma^2}\frac{1}{\left| \vect{r} -
  \vect{r^\prime} \right|} \, d\vect{r}^\prime = \frac{\erf(r/\sigma)}{r}~.
\end{equation}
An advantage of the Coulomb separation is that $\sigma$
can be chosen specifically in different applications so that condition i) is very well satisfied.

When all charges in the system (both fixed and mobile) are separated
using the same $\sigma$ and all other intermolecular interactions remain unchanged,
LMF theory then gives a mapping to a Coulomb mimic system where all
$1/r$ interactions are replaced by the short-ranged \vs\  and there is a
restructured electrostatic potential \Vr\  that satisfies the Coulomb LMF equation
\begin{equation}
  \Vr(\vect{r}) = \V_{0}(\vect{r}) +  \int d\vect{r}^\prime 
  \rhoq_{\rm R,tot}(\vect{r}^\prime ;[ \Vr ])\, v_1(\vdiff{r}{r^\prime}) + C.
\label{eqn:CoulombLMF}
\end{equation}
Here $\V_{0}(\vect{r})$ is the short-ranged part of the external potential, given by
the convolution of $\vs$ with the fixed charge
density, and $\rhoq_{\rm R,tot}(\vect{r}^\prime ;[ \Vr ])$
is the total equilibrium charge density
from both fixed and mobile charges.
An alternate form of Eq.\ (\ref{eqn:CoulombLMF}) better relates
LMF theory to conventional electrostatics \cite{Rodgers_Weeks2008}.
Noting the convolution defining $v_{1}$, we see that 
the slowly-varying part $\V_{\rm R1}(\vect{r}) \equiv \V_{\rm R}(\vect{r}) - \V_{0}(\vect{r})$
of the restructured potential in (\ref{eqn:CoulombLMF}) exactly satisfies Poisson's equation but
with a Gaussian-smoothed charge density $\rho^{q\sigma}_{\rm R,tot}$, given by the
convolution of $\rho^{q}_{\rm R,tot}$ with the Gaussian
in Eq.\ (\ref{eqn:v1def}).

We now apply the Coulomb LMF Eq.\ (\ref{eqn:CoulombLMF})
to the extended simple point charge (SPC/E) model for
water \cite{BerendsenGrigeraStraatsma.1987}.
As suggested by the SCA, we first consider
a Gaussian-truncated (GT) model, where all the $1/r$ interactions from charges in SPC/E water
are replaced by $v_{0}(r)$ and we ignore all structural effects from 
$v_{1}$. Bulk GT water with $\sigma = 0.45$ nm gives excellent results for both atom-atom
and dipole-angle correlation functions when compared to the
full SPC/E model, with Coulomb interactions treated by Ewald sums \cite{Rodgers_Weeks2008}. 

\begin{figure}[tbdp]
  \centering
    \includegraphics[height=3.0in,angle=270]{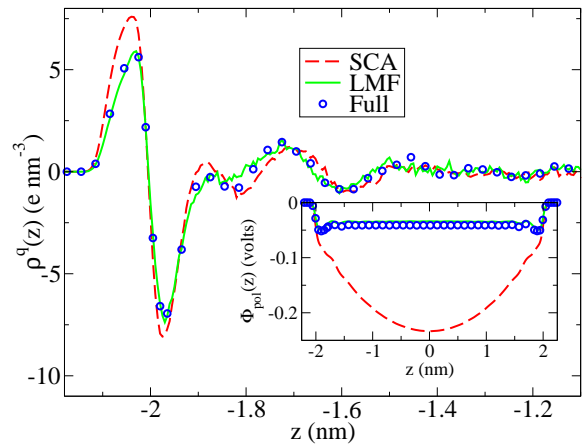}
\caption{Charge densities for water confined by hydrophobic LJ walls \cite{LEE:1984}
centered at $z= \pm 2.25$ nm
for the SCA (dashed line), mimic system (solid line) and
full system (circles).
The inset shows the corresponding polarization potentials \cite{sim-details}.}
\label{fig:waterchgdall}
\end{figure}

Moreover, as Fig.\ 2 shows, when GT water is
confined between two hydrophobic LJ walls as defined in \cite{LEE:1984}, the charge density
$\rho^{q}_{0}(z)$ determined by simulations \cite{sim-details} using bare wall fields
with $\V_{0} =0$ seems to capture most qualitative features of the dipole layer, a characteristic
property of water near extended hydrophobic interfaces \cite{Rodgers_Weeks2008,LEE:1984}. Only small
differences in the peak heights are visible when compared with full SPC/E water,
simulated using the slab-corrected Ewald 3D
method \cite{Yeh_Berkowitz1999}.

Nevertheless, as has long been recognized \cite{FellerPastorRojnuckari.1996,Spohr:1997},
major errors are seen in the polarization potential felt by a test charge, given by integrating Poisson's equation:
\begin{equation}
  \label{eqn:elecpotential}
  \Phi_{\text{pol}}(z) = -\int_{-L/2}^{z} dz^\prime \int_{-L/2}^{z^\prime} dz^{\prime\prime} \rho^{q}(z^{\prime\prime}).
\end{equation}
As shown in the inset of Fig.\ 2 for the full SPC/E model, $\Phi_{\text{pol}}$ should reach
a plateau in the central bulk region, and GT water fails dramatically in this respect.
\begin{figure}[tdp]
  \centering
    \includegraphics[height=3.0in,angle=270]{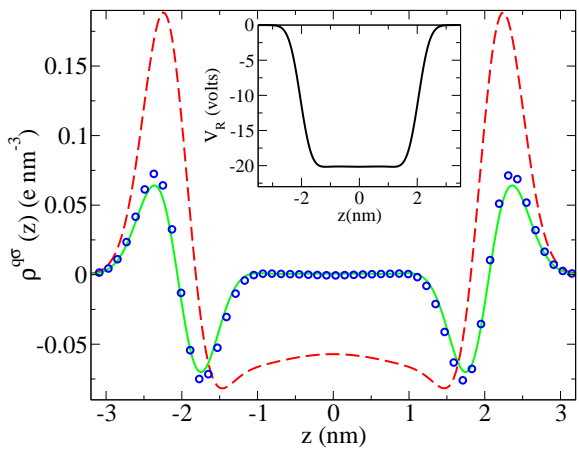}
\caption{Gaussian-smoothed charge densities with $\sigma =0.45$ nm. Labels are the same as in
Fig.~\ref{fig:waterchgdall}. The inset gives the effective
field ${\cal V}_{\rm R}({\mathbf r})$. Note that the bare ${\cal V}({\mathbf r})=0$ in this case.}
\end{figure}

A self-consistent solution of Eq.\ (\ref{eqn:CoulombLMF}) yields a very accurate charge
density that corrects all such failures.
The inset in Fig.\ 3 gives the converged $\V_{\rm R}(z)$.
Gaussian smoothing of the bare charge density as dictated by LMF theory
averages over the simulation noise and local structure
and reveals the much smaller coherent long-wavelength features
that control the electrostatics \cite{Rodgers_Weeks2008}.
The smoothed charge density quickly decays to a neutral bulk in Fig.\ 3 for both
SPC/E water and LMF theory, while a true bulk never forms in GT water,
causing the very poor polarization potential in Fig.\ 2.


These results highlight the advantages of the new LR-LMF method.
It reduced the simulation time by more than an order of magnitude compared to
use of the standard iteration method  \cite{Rodgers_Weeks2008}
or to use of the slab-corrected Ewald 3D method \cite{Yeh_Berkowitz1999}.
When using the standard iteration method,
it proved very difficult to distinguish between
equilibrium charge density fluctuations, present for any given field,
and the desired changes in the charge density
as the iteration proceeded to self-consistency. To obtain convergence,
earlier workers had to use a large value of $\sigma = 0.6$ nm and
the charge density at each iteration was taken as the average over a set of 10 parallel simulations
with different initial conditions \cite{Rodgers_Weeks2008}.

In contrast, the results in Figs.\ 2 and 3
were obtained from only two very short trial simulations \cite{sim-details},
starting first with the SCA $\V_{0} =0$, and using the smaller bulk $\sigma = 4.5$ nm.  As before,
the EXP form (\ref{eqn:rhoexact}) was very noisy when used with SCA configurations,
but the LR form (\ref{eqn:linearPhi}) was much better behaved.
It generated a second trial field in the linear regime that
was virtually identical to the final self-consistent field shown in the inset of Fig.\ (3).
The remaining effects of equilibrium fluctuations in the mimic system show
up as small variations (about the size of the circle symbol) in the bulk value of the polarization potential
in Fig.\ 2 or the heights of the charge density peaks in Fig.\ 3.
As shown in the supplementary material,
excellent results have also been obtained for
ion pairing in ionic solution models using the LR-LMF method \cite{sim-details}.

Reaction field (RF) truncations of Coulomb interactions
have recently been used in massively-parallel
simulations of biological systems to permit
much faster simulations \cite{Schulz2009}. 
LR-LMF theory may provide
a promising linear-scaling alternative in which the effective field corrects
known problems like those illustrated in Fig.\ 2
arising from simple RF or SCA truncations. 
More generally, we believe the  efficient solutions generated
by the LR-LMF method establish the full power of the
basic mean field picture for both quantitative and qualitative 
analysis of a wide range of electrostatic and dielectric
phenomena in nonuniform liquids.

This work was supported by the National Science Foundation (grants CHE0628178 and CHE0848574).
We are grateful to Gerhard Hummer, Chris Jarzynski, and Jocelyn Rodgers
for very helpful remarks.


\begin{thebibliography}{15}
\bibitem{generalfield} A general field
$\phi(\vect{r})$ can be used in Eq.\ (\ref{eqn:LMFGeneral}) as well.
\end{thebibliography}

\end{document}


\title{Supplementary Material for:\\
Efficient solutions of self-consistent mean field equations for dewetting and electrostatics in nonuniform liquids}

\author{Zhonghan Hu}
\altaffiliation[Present address: ]{State Key Laboratory of Supramolecular Structure and Materials, Jilin
University, Changchun 130012, China}

\author{John D. Weeks}
\affiliation{Institute for Physical Science and Technology and
Department of Chemistry and Biochemistry, University of
Maryland, College Park, Maryland 20742}

\maketitle

\section{Simulation details for dewetting of the LJ fluid around a hard sphere solute and
for water confined between hydrophobic walls}
In the following, equation and figure numbers refer to the main paper, unless otherwise indicated.
Our Monte Carlo (MC) simulation for $N=2363$ LJ particles with a hard sphere solute at a state near the triple point
with $T=0.85$ and $\rho^B=0.70$
follows the work of Huang and Chandler \cite{Huang_Chandler2000}.
Configurations of a short trial simulation of 10000 MC steps
based on the SCA and Eq.\ (4) were used in the LMF Eq.\ (1) to generate the trial field given by the solid line in Fig. (1b).
A second trial simulation of 20000 steps using that field generated the final self-consistent LMF.
Both runs were much shorter than the final production run of 200000 steps needed to determine the
smooth profile of the full LJ system given by the circles in Fig.\ (1a).

Our molecular dynamics (MD) simulation of SPC/E water confined
between hydrophobic LJ walls used the same system parameters
and a modified {\sc dlpoly2.18} MD
package~\cite{SmithYongRodger.2002.DLPOLY:-Application-to-molecular-simulation}
similar to that described in detail in \cite{Rodgers_Weeks2008a}, with further modifications
to save configurations needed for the LR-LMF theory.  
Two trial simulations with $\sigma=4.5$ nm each of 1 ns length and starting from the SCA
were used to generate the data in Figs. (2)
and (3).  This was more than an order of magnitude faster than analogous results
obtained using the standard iteration method \cite{Rodgers_Weeks2008a}.
Moreover, as discussed in detail in \cite{Rodgersthesis}, a large value of $\sigma = 0.6$ nm
had to be used in the standard iteration method, because of problems in distinguishing
equilibrium charge density fluctuations with a given estimate for the effective field
from the desired iterative changes in the charge density.
Even after averaging over results of 10 parallel simulations for each iteration, convergence
was not found with $\sigma=4.5$ nm.   

In contrast, the LR-LMF method uses the same set of
equilibrium trial system configurations to determine the remaining changes in the charge density
as self-consistency is achieved by iterating the LMF equation.
Convergence was found using the same $\sigma=4.5$ nm used for the bulk, and required
only two short trial simulations. 


\section{case study of dilute ionic system}

To further illustrate the versatility of the LR-LMF method we discuss here results for a
dilute bulk ionic system at moderately strong
coupling. This presents a challenge to theory, since the coupling is strong enough that traditional methods 
based on Poisson-Boltzmann theory will fail, but weak enough that the SCA alone will not be
accurate unless a very large $\sigma$ is used. Indeed, previous applications of LMF theory to bulk
ionic solution models considered only strong-coupling states and used such a large
$\sigma$ that the SCA alone was accurate \cite{Chen_Weeks2006}. The present
calculation represents the first 
self-consistent solution of the full LMF equation for a general ionic solution model in this
difficult regime. As we will see only a single simulation
in this case is needed to achieve full self-consistency.

The system is a size-asymmetric primitive model consisting of charged LJ molecules.
The LJ interaction parameters are $\sigma_+=4.0 {\rm \AA}$,
$\sigma_-=3.0{\rm \AA}$, $\sigma_{+-}=3.5{\rm \AA}$, $\varepsilon_+=0.1
{\rm kcal/mol}$, $\varepsilon_-=0.4 {\rm kcal/mol}$ and
$\varepsilon_{+-}=0.2 {\rm kcal/mol}$. 
The cation and anion has unit charge and mass of 12 and 30 respectively.
The cutoff distance for both the LJ and Ewald short ranged interaction in the system is 15 {\rm \AA}. 
The system has 1024 pairs of ions in a cubic box of size $62{\rm \AA}$ with $T=5000$ K.
The MD simulation of the full system using Ewald sums used 90000 steps.

\begin{figure}[htdp]
 \centering
    \includegraphics[height=3.0in,angle=270]{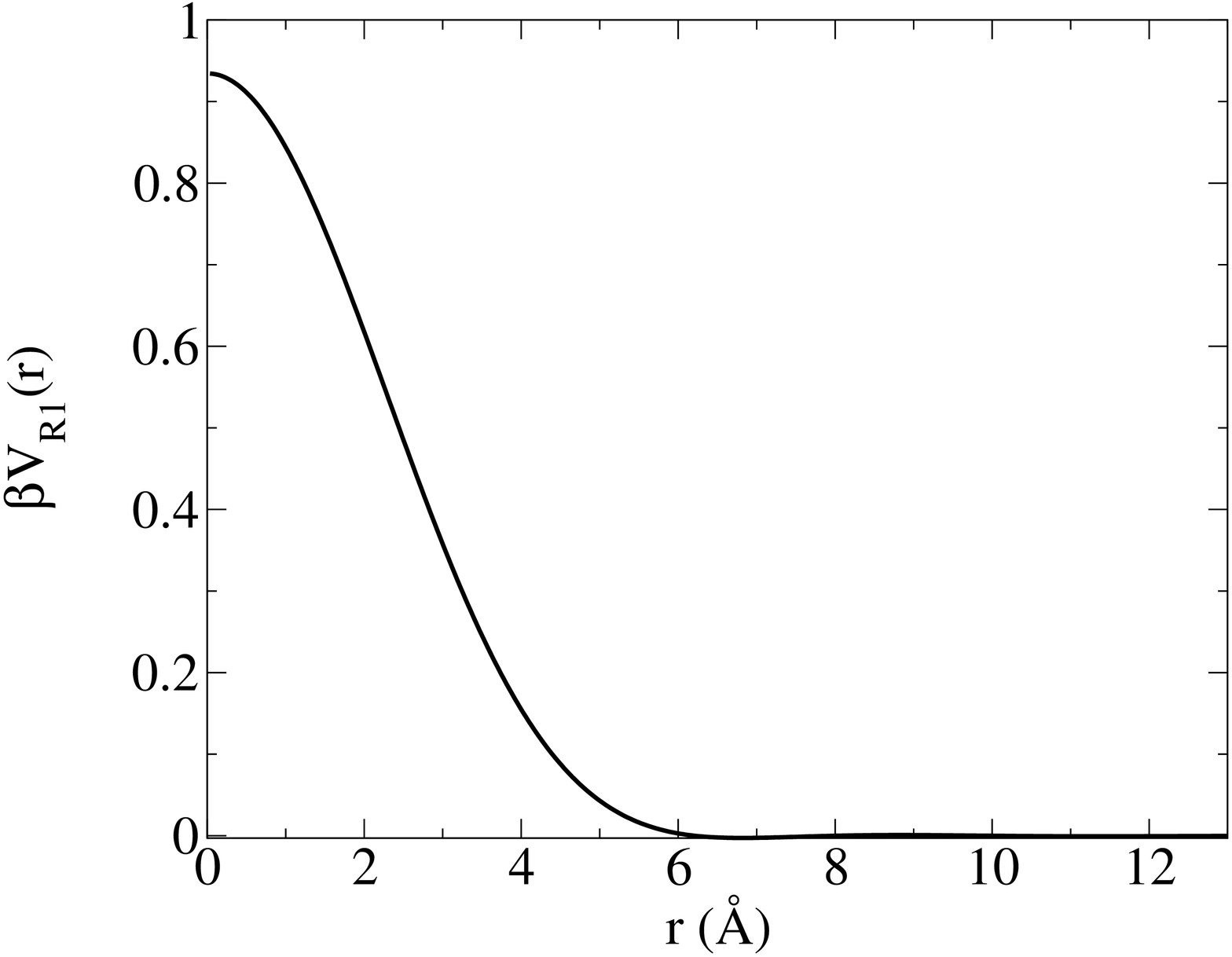}
\caption{The restructured field with a cation fixed at the origin}
\end{figure}
 
\begin{figure}[tdp]
 \centering
    \includegraphics[height=3.0in,angle=270]{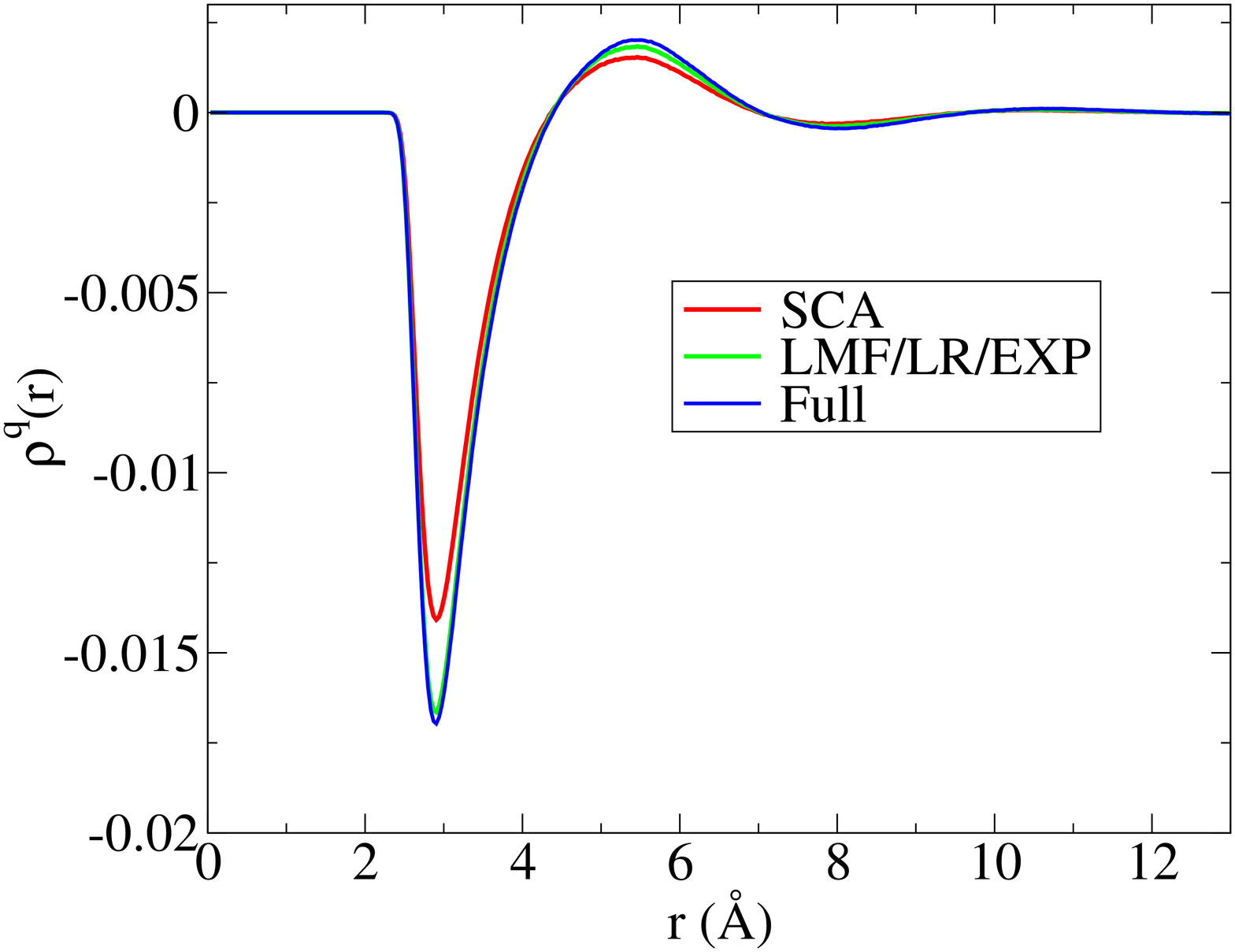}
\caption{The charge density around a cation}
\end{figure} 

Results of MD simulations for the full system and for the LR-LMF method using the SCA as the
trial system are presented here.
Fig. (1) shows the restructured field with a cation fixed at the origin.
The charge density around the cation is shown in Fig. (2).
LR-LMF and EXP-LMF results overlap starting from SCA configurations
and we show only the former in the graphs.
Significant ion pairing is indicated by the strong first peak in the cation-anion correlation function in Fig. (3).
The SCA simulation used $\sigma = 3.5{\rm \AA}$, a value for which the
SCA correlation function had noticeable errors, as seen in Fig. (3).
However the trial system based on SCA configurations is
in the linear regime, and no further simulations are needed to get the
converged results shown here from iteration of the LMF equation. 
  
\begin{figure}[tdp]
 \centering
    \includegraphics[height=3.0in,angle=270]{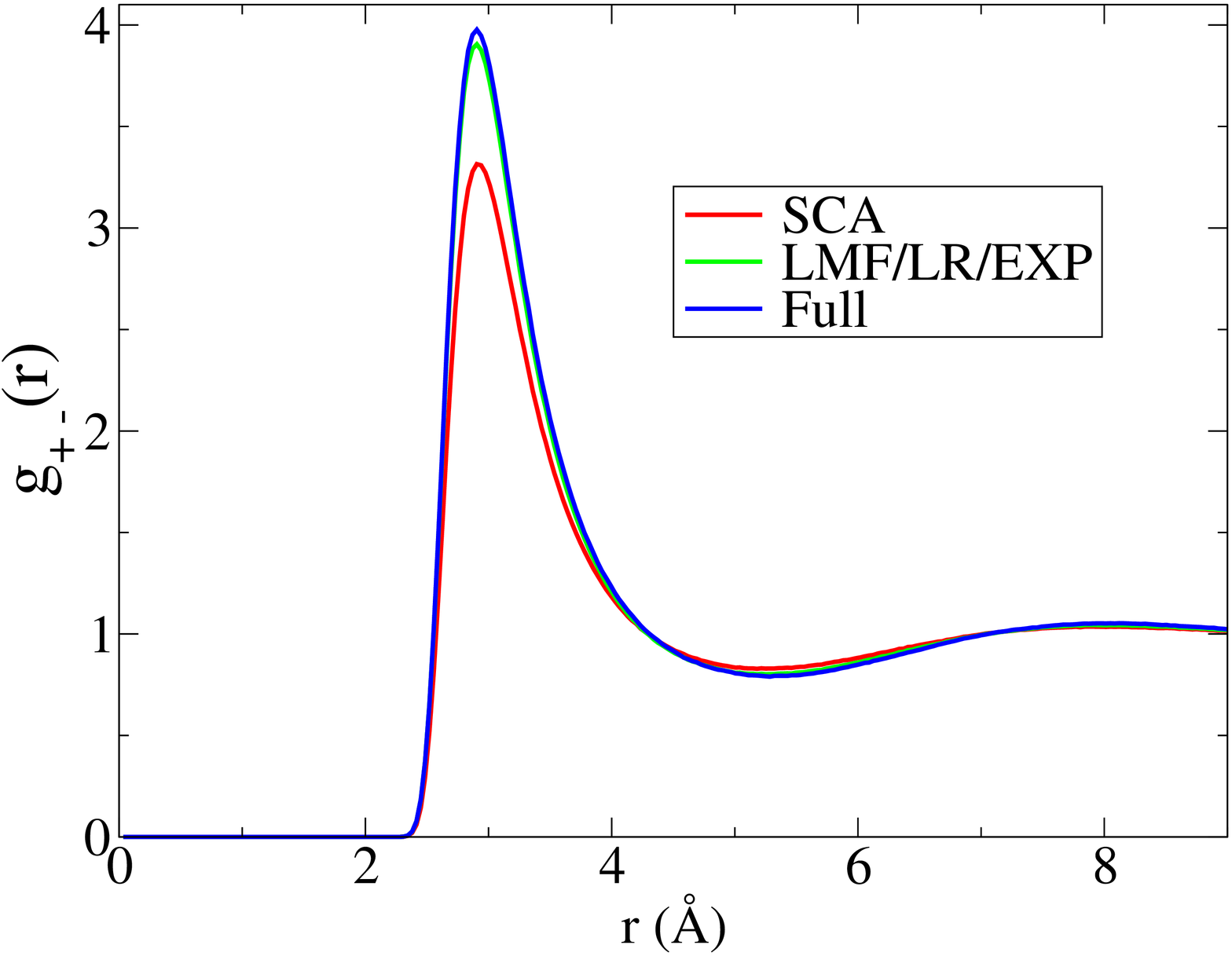}
\caption{The cation-anion correlation function}
\end{figure}

As explained in \cite{Chen_Weeks2004a}, it is useful to rewrite the LMF equation in a more
numerically stable form for iteration so that small errors in the decay of the charge density are not amplified when
convoluted with the long-ranged $v_{1}$. We used the specific method described in
Appendix E of \cite{Rodgersthesis}.

It is gratifying that LR-LMF theory allows for such accurate results using
a small value of $\sigma$, since this permits fast mimic system simulations. 
However, care must be taken, since the underlying LMF theory itself will fail if too small a $\sigma$ is used,
and the small differences still visible for the LR-LMF and full correlation functions suggests we are
close to the minimum accurate value.